# INCREASING ENERGY EFFICIENCY IN SENSOR NETWORKS: BLUE NOISE SAMPLING AND NON-CONVEX MATRIX COMPLETION

Angshul Majumdar and Rabab K. Ward
{angshulm and rababw}@ece.ubc.ca
Department of Electrical and Computer Engineering, University of British Columbia

Abstract –

The energy cost of a sensor network is dominated by the data acquisition and communication cost of individual sensors. At each sampling instant it is unnecessary to sample and communicate the data at all sensors since the data is highly redundant. We find that, if only a random subset of the sensors acquires and transmits the sample values, it is possible to estimate the sample values at all the sensors under certain realistic assumptions. Since only a subset of all the sensors is active at each sampling instant, the energy cost of the network is reduced over time. When the sensor nodes are assumed to lie on a regular rectangular grid, the problem can be recast as a low-rank matrix completion problem. Current theoretical work on matrix completion relies on purely random sampling strategies and convex estimation algorithms. In this work, we will empirically show that better reconstruction results are obtained when more sophisticated sampling schemes are used followed by non-convex matrix completion algorithms. We find that the proposed approach coupling blue-noise sampling with non-convex reconstruction algorithm, gives surprisingly good results.

1. INTRODUCTION

Probably the most pressing issue in distributed wireless sensor networks is the problem of energy efficiency. Sensor nodes are deployed in locations, where it is not feasible to replace its power source. Therefore the longevity of the sensor network is inversely proportional to the power consumption of each sensor. Reducing the energy consumption in sensor networks is an active area of research, and a recent comprehensive survey on this topic can be found in [1]. In this work, we propose a completely new approach to increase the energy efficiency of sensor networks.

Sensors consists of four modules – 1) Data acquisition module for sensing the data from the environment; 2) Processing module for quantizing and encoding the data and making it fit for transmission; 3) Communication module for receiving/transmitting information/data to-and-from the base station (sink); 4) Battery module to serve as the seat of power for the other modules. Of the first three modules, the data acquisition and communication modules are the most power hungry. The processing module consumes negligible amount (1000 times less) of power compared to the other two. Therefore while proposing methods for energy efficient sensor networks one must keep in mind to reduce the operation of the data acquisition and/or the communication module.

In this work, we propose a new method to reduce the energy consumption of the sensor network. It will reduce the average number of operations for both the data acquisition and the communication modules. The idea is a simple one. Assume that the sensor nodes lie on a regular rectangular grid. At each sampling instant, only a subset of all the nodes samples and transmits the data to the sink. There is a smart algorithm at the sink, that estimates the data at all (sampled + unsampled) the sensor nodes.

The questions that needs to be answered now are – how to select the subset of the sensor nodes at each sampling instant and how to reconstruct the data in from the partially sampled data. Answering these two questions is the focus of this work.

For the time being let us assume that the sampling strategy and the reconstruction algorithm is known. We now analyze, why such a proposed method will lead to an energy efficient sensor network. At each sampling instant only a small subset (say 20%) of all the sensors needs to acquire and transmit the data. At the next instant a different subset of sensors will perform the same operations. Thus on an average the proposed scheme reduces the energy consumption of the sensors by five times; in other words the life of the sensor network is increased five-fold!

Since we assume that the sensors lie on a regular rectangular grid, the said problem can be recast as a matrix completion problem [2-4]. The sensors lie on a uniform rectangular grid. The sample values at all the sensors therefore form a matrix. In the proposed approach only a subset of the samples are available, i.e, the matrix is partially sampled. The problem is to reconstruct the entire matrix from these partially sampled entries. Therefore the problem is that of a matrix completion.

Matrix completion is a recent active area in applied mathematics. The theoretical results have been proved for random sampling schemes followed by a convex optimization (Nuclear norm minimization) algorithm for reconstruction. In this work, we will empirically extend the matrix completion problem in two ways – 1) employing more sophisticated blue-noise sampling schemes like Quasi-random sampling, Quasi-crystal sampling and Farthest Point Sampling; 2) employing powerful non-convex optimization based reconstruction algorithms.

The rest of the paper is organized into several sections. The following section contains the theoretical background of matrix completion. It also contains discussions on the closely related topic of Compressed Sensing [5, 6]. In Section 3, we discuss the proposed approach in detail. Section 4, discusses the different sampling strategies. Section 5, develops Majorization-Minimization framework for convex and non-convex optimization algorithms for data reconstruction. Experimental results are shown in Section 6. The conclusions of this work and future directions of research are finally discussed in Section 7.

2. MATRIX COMPLETION THEORY

There is a matrix $X_{n \times n}$ (it can be rectangular as well). But all the entries ($x_{i,j}, (i,j)=1...n$) are not available. Only a subset ($\Omega$) of entries are observed. Now the question is – is it possible to estimate all the entries of the matrix given the set of partially observed samples? In general, the answer is NO. However, in a special situation, when the matrix is of low rank, it is possible to estimate the entire matrix provided 'enough' samples are available.

The intuitive reason behind this argument is that, if the rank of the matrix $r < n$ is low, then the number of degrees of freedom is only $r(2n-r)$. When the rank ($r$) is small, the degrees of freedom is considerably less than the total number of entries ($n \times n$) in the matrix. Therefore, we can hope there might be a possibility that all the entries of the matrix are recoverable from a subset of it.

The above discussion can be framed in the following optimization problem,

$$\min \ rank(X)$$
$$\text{subject to } Y = M_\Omega(X) \tag{1}$$

where, $M_\Omega$ is the masking operator that selects the entries in $X$ falling in the set $\Omega$, $Y$ are the available samples.

Unfortunately (1) is an NP hard problem; the complexity of solving this problem is doubly exponential!

Researchers in machine learning and signal processing has been interested in this problem in the past few years for a variety of problems – Collaborative Filtering, System Identification, Global Positioning, Direction of Arrival estimation etc. What they did was to solve the most plausible (convex) surrogate to (1), which is the following,

$$\min \ \|X\|_* \quad \text{subject to } Y = M_\Omega(X) \tag{2}$$

where $\|.\|_*$ is the nuclear norm (also called Trace norm or Ky-Fan norm) of the matrix, and is defined as the sum of absolute singular values of the matrix.

The nuclear norm is the closest (tightest) convex relaxation to the NP hard rank minimization problem. Therefore, solving (2) and expecting a result similar to (1) was in some way expected.

There had been a number of works in machine learning that was related to the solution of (2). In [7] the nuclear norm minimization problem was solved for the purpose of system identification. A variant of (2) was employed to solve the problem of multi-task classification in [8]. In [9] the consistency of the optimization problem (2) was discussed.

In recent times, mathematicians are taking a closer look at the matrix completion problem and its relation with nuclear norm minimization. They are mainly interested in finding the bounds on the number of entries required to estimate the complete matrix and the conditions necessary to be imposed on the nature of the matrix and the mask (sampling strategy). We will briefly discuss these theoretical results.

The fundamental theorem behind the matrix completion problem states that –

Suppose we observe $m$ entries of $X$ with locations sampled uniformly at random. Then there is a positive numerical constant $C$ such that if

$$m \geq C\mu^a nr \log^b n \tag{3}$$

the nuclear norm minimization problem (2) will find the unique solution with probability $1 - n^{-d}$.

Here, $\mu$ depends on the nature of $X$ (will be discussed shortly).

The integer constants a, b and d vary depending on the approach of the proof. What is important to note that the theorem proves that an NP hard problem (1) can be solved by a convex relaxation (2) with somewhat larger number of samples with a very high probability.

Now consider a pathological low rank matrix which has all entries as zeros except for a single non-zero value. Even though the matrix is of rank 1, it is impossible to recover the matrix without sampling all the entries. The same is true for a matrix which has only one row/column of non-zero entries while the rest are zeros. For other examples of such pathological matrices, see [2-4]. The behavior of such matrices can be explained by the understanding the factor $\mu$.

Let $X = U\Sigma V^T$ be the SVD of the matrix under consideration; where $U = [u_1, ..., u_r]$, $V = [v_1, ..., v_r]$ be the right and left singular vectors respectively and $\Sigma = diag(\sigma_1, ..., \sigma_r)$ be the singular values. Now $\mu$ is defined as,

$$\max(u_k) \leq \sqrt{\mu/n} \text{ and } \max(v_k) \leq \sqrt{\mu/n} \tag{4}$$

where max(*w*) is the maximum absolute value in the vector *w*.

Expressed in words, the singular vectors should not be too spiky. If they are spiky, then the $\mu$ is high, and therefore the number of samples needed for perfect recovery (3) is also high. This is called the 'Incoherence Property'. For the pathological matrices discussed earlier, the singular values are spiky and therefore would require lot of samples to estimate the matrix.

Lastly we will discuss why the theorem has been proved for samples 'collected uniformly at random'. If there are no entries selected from a particular row or column, then it is impossible to reconstruct that row or column for even a matrix of rank unity. When the samples are collected uniformly at random, the number of samples required to ensure every row and column is sampled at least once is $O(n\log n)$; this is as good an estimate one can get for a problem of size $n^2$.

### 2.1 Connections with Compressed Sensing

Compressed sensing studies the problem of solving a system of under-determined linear equations when the solution is known to be sparse, i.e. consider the following system of equations,

$$y_{m\times 1} = M_{m\times n} x_{n\times 1}, \; m < n \tag{5}$$

In general, (5) has infinitely many solutions. But if the solution is known to be sparse it has been proved [10] that the solution is necessarily unique. Assume that the vector *x* is k-sparse, i.e. it has *k* non-zero entries while the rest are all zeroes. In such a scenario, there are only *2k* (k-positions and k-values) unknowns. Now as long as the number of equations $m \geq 2k$, it is possible to solve (5). Mathematically, the problem can be stated as,

$$\begin{aligned} &\min \| x \|_0 \\ &\text{subject to } y = Mx \end{aligned} \tag{6}$$

$\| . \|_0$ is not a norm in the strictest sense, it only counts the number of non-zero entries in the vector.

In words (6) means, that of all the possible solutions (5), chose the sparsest one.

Unfortunately solving (6) is known to be an NP hard problem [11]. There is no known algorithm which has shown any significant improvement compared to brute force solution of (5). In machine learning [12] and signal processing [13], instead of solving the NP hard $l_0$-norm minimization problem, its closest convex surrogate ($l_1$-norm minimization problem) is generally solved instead,

$$\begin{aligned} &\min \| x \|_1 \\ &\text{subject to } y = Mx \end{aligned} \tag{7}$$

In [10], it was shown that for several types of matrices (*M*) solving (6) and (7) was equivalent, i.e. both of them yielded the sparsest solution.

Solving (7) is easy since it is a convex problem and can be solved by linear programming. But, the number of equations required to solve (5) via convex optimization (7) is significantly larger when compared to solving the NP hard problem (6). The number of equations required also depends on the type of matrix (*M*); for some common matrices the number of equations required is [14],

$$m \geq Ck \log(n) \tag{8}$$

The trade-off is expected; the ease of solving the inverse problem comes at the cost of larger number of equations!

Both CS and MC study problems where the number of unknowns is apparently larger than the number of equations. In CS the length of the vector to be estimated is larger than the number of equations; in MC only a subset of the elements of the matrix are known. In general, none of the problems have a unique solution. However, when the degrees of freedom in the solution are less than the number of equations, both problems have a unique solution. For CS the degree of freedom is *2k*, where *k* is the number of non-zero elements in the vector; for MC the degree of freedom is *r(2n-r)*, where *r* is the rank of the matrix and $n^2$ is the number of elements in the matrix.

The original problem to be solved in both CS and MC are NP hard. Fortunately, it has been proved that instead of solving the NP hard problems (1) and (6), convex surrogates (2) and (7) can be solved instead. The cost to be paid for solving the easy convex problem instead of the NP hard problem is an increase in the number of equations (3), (8). There are standard packages for solving (2) and (7); (2) can be solved by semi-definite programming (SDP) while (7) can be solved by linear programming (LP). Standard solvers are available both for SDP and LP. Unfortunately, such packages are very slow. In CS, developing specialized fast optimization algorithms for (7) such as SPGL1 [15], C-SALSA [16] and NESTA [17], is an active area of research. A similar trend is observed for MC as well. There are several specialized algorithms to solve the MC problem [18-20]; how the matrix completion algorithms are not derived with the same rigour as their compressed sensing counterparts.

In CS the NP hard problem (6) and the convex problem (7) are two extremes. The first one is hard to solve but requires very few equations, while the latter is easy to solve but requires considerably more equations. There is a solution that lies between (6) and (7); it is the so called fractional norm minimization problem.

$$\min \|x\|_p \quad 0 < p \leq 1$$
$$\text{subject to } y = Mx \tag{8}$$

The problem (8) is non-convex. It is not difficult to solve, but is not guaranteed to converge to the global minimum. However, practically it has been found to give extremely good results [21-22].

It has been shown in [22], that the number of equations required for solving (8) is,

$$m = C_1 k + p C_2 k \log(n) \tag{9}$$

When the value of fractional norm p is small, the second term almost vanishes and the number of equations needed is only in multiples of the number of non-zero elements *k*.

In this work we are motivated by the findings of non-convex compressed sensing algorithms for solving (9). Instead of solving (2), we expect perfect estimation of the matrix X, with fewer entries if the following non-convex problem is solved,

$$\min \|X\|_{p*}$$
$$\text{subject to } Y = M_\Omega(X) \tag{10}$$

where $\|X\|_{p*} = (\sum_{i=1}^{r} \sigma_i^p)^{1/p}$, $0 < p \leq 1$, $X = U\Sigma V^T$ and $\Sigma = diag(\sigma_i)$.

We just change the objective function. Earlier it was convex, being the sum ($l_1$-norm) of singular values, now it is non-convex since it is an $l_p$-norm of the singular values.

## 3. PROPOSED APPROACH

In this work we assume that the sensors are arranged in a regular rectangular grid. This assumption may not be exactly satisfied in practice, but a recent work exploiting compressed sensing techniques for data reconstruction in wireless sensor network made the same assumption [32]. At each sampling instant only a subset of all the nodes is active. These nodes acquire/sample the data at that instant and communicate it to the sink. The sink, employs a matrix completion algorithm to estimate the data for all the nodes.

We have to show that sensor network data indeed satisfies the properties required by the theory of matrix completion. For the experiments, we did not have access to real sensor network data. Therefore we generated the data from a standard tool [33]. For more discussion on this toolbox we request the reader to peruse [34].

The tool generated data of size 64X64. The generated data is spatially correlated. The correlations can be varied by changing different parameters in the tool. In this paper, we worked with three levels of correlation – Low, Medium and High (the parameters for these three levels of correlations were suggested in the tool itself). Our first observation is – for all three different levels of correlation, the generated data matrix has a low rank. The following figure corroborates our observation.

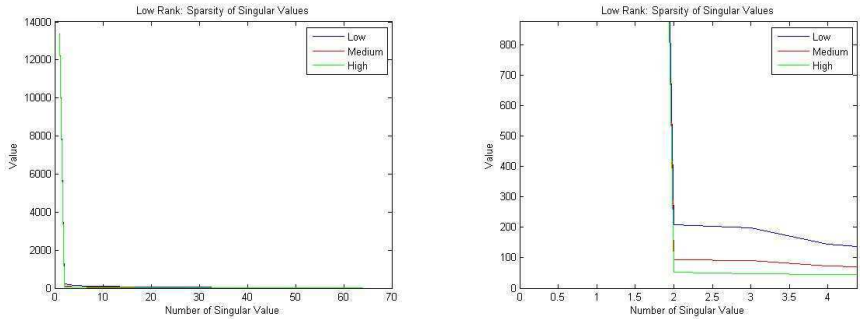

Fig. 1. Left: Sorted Singular Values; Right: Enlarged

Fig. 1. shows that the matrices indeed have a very low rank. We calculated what fraction of the total energy is captured in the top singular values. For all the three different levels of correlation, the first six singular values captured more than 99.97% of the total energy! Thus, we see that the simulated sensor network matrices are indeed of low rank. Therefore we can hope to apply matrix completion algorithms to estimate the full data. But we are not there yet. One still needs to check the incoherence property of the right and left singular vectors for the generated data.

Table 1. Coherence Measure for Singular Vectors

| Correlation | max($u_k$) – left singular vectors | max($v_k$) – right singular vectors | μ - Coherence |
|---|---|---|---|
| Low | 0.7592 | 0.4603 | 37 |
| Medium | 0.7276 | 0.4944 | 34 |
| High | 0.4454 | 0.4491 | 13 |

The incoherence measure is of $O(2)$.

The last condition, the matrix completion theory requires is that the collected samples should not completely miss a row or a column. Theorists have proven the results for uniform random sampling, since this kind of sampling satisfies the said requirement up to a logarithmic factor. However, our work is inspired by recent findings in compressed sensing [23], where it is empirically seen that better reconstruction results can be obtained when blue noise sampling strategies are employed instead of uniform random sampling.

In the previous section, we have mentioned that prevalent algorithms in matrix completion rely on convex optimization. In this work, we are motivated by findings in non-convex compressed sensing. Non-convex optimization algorithms in compressed sensing have shown that better reconstruction results can be obtained compared to convex optimization based algorithms. In this work, we propose a novel non-convex matrix completion algorithm to achieve better reconstruction.

4. SAMPLING STRATEGIES

In [2-4], it the number of entries (3) required for perfect estimation of the rank deficient matrix was based on the assumption that the samples from the matrix have been selected uniformly at random. Ideally one needs a sampling strategy which does not leave completely omit any row or column of the matrix. When sampling is performed uniformly at random, the number of samples needed to ensure this condition is $O(n\log n)$. This intuitively explains the logarithmic factor in (3). However uniform random sampling may not yield the best achievable results.

It has been found that while applying compressed sensing techniques to seismic [23] and magnetic resonance [24] imaging, better results can be obtained if blue noise sampling schemes like farthest point sampling and poisson disk sampling, are used instead of uniform random sampling. Blue noise sampling schemes are generally employed in computer graphics; there are no theoretical results in signal processing regarding their optimality in signal reconstruction. However, in this work, we are inspired by the empirical results in compressed sensing, and investigate the performance of these blue noise sampling strategies for matrix completion technique when applied to the distributed sensor network problem.

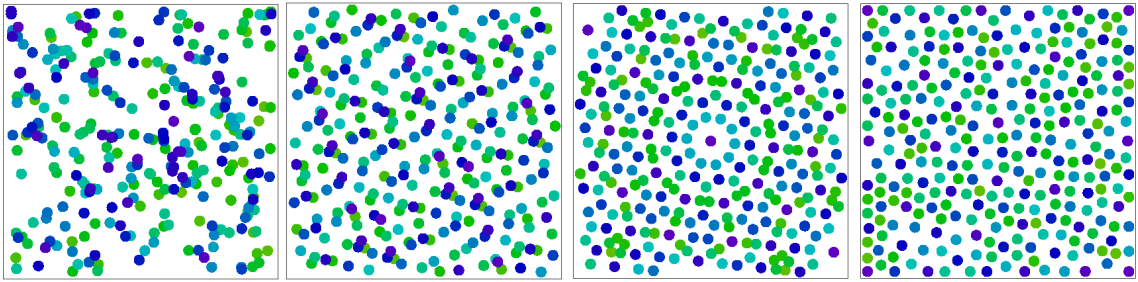

Figure 2. From Left to Right – Random, Quasi-random, Quasi-crystal and Farthest Point [25]

Random Sampling – It aims at global variability. It can be implemented either from a 2D uniform random distribution or as a random walk on a unit square. It is the most popular sampling method in signal reconstruction literature.

Quasi-random Sampling – It is similar to random sampling but is less cluttered and more evenly distributed. It can be efficiently generated by a Halton or Sobol sequence. It has been used sometimes for signal reconstruction problems [26] mainly as an alternatively to Shannon sampling.

Quasi-crystal Sampling – It aims at local regularity. It is not a popular choice owing to its high computational requirement, it has been shown [27] that they are more optimal than random sampling for signal reconstruction problems.

Farthest Point Sampling – It aims at spatial uniformity. The main idea behind this is to repeatedly place the next sample point in the middle of the least known area of the sampling domain. In this work, fast marching algorithm is used to generate this sampling.

5. RECONSTRUCTION ALGORITHMS

Solving (2) requires semi-definite programming (SDP). However, SDP is slow. Therefore in recent times specialized algorithms for solving the matrix completion problem has been proposed [17-19]. The proposed algorithms show good results but are not based on any rigorous optimization method. One contribution of this work will be to develop matrix completion algorithms from a more rigorous footing. In this work, we particularly follow the Majorization-Minimization (MM) technique for solving the optimization problems.

The optimization problems (2) and (10) have equality constraints. However in practical sensor networks, the measurements will be corrupted by noise. Therefore a more practical data acquisition model will be,

$$Y = M_\Omega(X) + \eta \tag{11}$$

The noise is assumed to be distributed Normally with zero mean.

In the noisy scenario, instead of solving the optimization problem with equality constraints (2) and (10), we would like to solve them with a quadratic mismatch constraint. The problem takes the following form,

$$\min \|X\|_{p*}$$
$$\text{subject to } \|Y - M_\Omega(X)\|_F^2 \leq \varepsilon \tag{12}$$

where $\|.\|_F$ is the Frobenius norm of the matrix.

When p = 1, the problem is convex, and for fractional values of p, it is non-convex. Theoretical analysis for the convex case is discussed in [3].

The constrained optimization problem (12) is hard. There is an easier unconstrained equivalent of (12) which takes the form,

$$\min \|Y - M_\Omega(X)\|_F^2 + \lambda \|X\|_{p*} \tag{13}$$

The form (13) is somewhat easier to solve. The parameters $\lambda$ and $\varepsilon$ are related; however for all the interesting problems it is difficult to find the relation analytically. In this work, we develop an optimization algorithm to solve (13). Later, we will show how to solve (12) via (13) algorithmically.

a. Majorization – Minimization

A more convenient way to represent (13) is the following,

$$\min \ \|y - Mx\|_2^2 + \lambda \|X\|_{p*} \tag{14}$$

where $x_{n^2 \times 1}$ is the vectorized version of the matrix $X_{n \times n}$ formed by row/concatenation; $M: \Re^{n^2} \to \Re^m$ is a restriction operator, which has diagonal elements as ones at the sampling locations; $y_{m \times 1}$ is the vector of measurements. The notation may be slightly non-standard but we are consistent with this notation throughout the text.

We follow the Majorization-minimization (MM) procedure outlined in [28]. The problem to minimize is,

$$J(x) = \|y - Mx\|_2^2 + \lambda \|X\|_{p*} \tag{15}$$

MM replaces the hard minimization problem $J(x)$ by an iteration of easy minimization problems $G_k(x)$. The iterations produce a series of estimates which converge to the desired solution, i.e. the minimum of (15).

MM Algorithm

Initialize: iteration counter k = 0; initial estimate $x_0$.

Repeat the following steps until a suitable exit criterion is met.

1. Chose $G_k(x)$ such that:
   1.1. $G_k(x) \geq J(x), \forall x$
   1.2. $G_k(x_k) = J(x_k)$
2. Set: $x_{k+1} = \min G_k(x)$
3. Set: k=k+1 and return to step 1.

The MM approach is popular in CS literature [29]. In this work, we will extend the same methodology to the problem of matrix completion.

    b. Landweber Iterations

First let us consider the minimization of the following optimization problem,

$$J(x) = \|y - Mx\|_2^2$$

For this minimization, $G_k(x)$ is chosen to be,

$$G_k(x) = \|y - Mx\|_2^2 + (x - x_k)^T (\alpha I - M^T M)(x - x_k) \tag{16}$$

where $\alpha$ is the maximum eigenvalue of the matrix $M^T M$. For our problem, when $M$ is a restriction operator $\alpha$ is unity.

Now minimizing $G_k(x)$ we get the famous Landweber iterations,

$$x_{k+1} = x_k + \frac{1}{\alpha} M^T (y - M x_k) \tag{17}$$

Using this update formula, $G_k(x)$ can be expressed as,

$$G_k(x) = \alpha \| x - x_{k+1} \|_2^2 - \alpha x_{k+1}^T x_{k+1} + y^T y + x_k^T (\alpha I - M^T M) x_k \tag{18}$$

Note that all the terms apart from the first term are independent of $x$ and does not play any part in the minimization.

### c. Nuclear Norm Minimization via Shrinkage

The problem to be solved is,

$$J(x) = \| y - Mx \|_2^2 + \lambda \| X \|_*$$

After the discussion in the previous sub-section the choice of $G_k(x)$ is quite obvious,

$$G_k(x) = \alpha \| x - x_{k+1} \|_2^2 - \alpha x_{k+1}^T x_{k+1} + y^T y + x_k^T (\alpha I - M^T M) x_k + \lambda \| X \|_* \tag{19}$$

Now minimizing (19) is the same as minimizing the following,

$$G'_k(x) = \| x - x_{k+1} \|_2^2 + \frac{\lambda}{\alpha} \| X \|_* \tag{20}$$

Since the other terms are independent of $x$.

Now, $x$ and $x_{k+1}$ are vectorized forms of matrices. The following property of singular value decomposition holds in general,

$$\| A_1 - A_2 \|_F^2 \geq \| \Sigma_1 - \Sigma_2 \|_F^2 \tag{21}$$

where $A_1$ and $A_2$ denotes two matrices and $\Sigma_1$ and $\Sigma_2$ are their singular value matrices respectively.

Using this property, minimizing (20) is the same as minimizing the following,

$$G''_k(s) = \| s - s_{k+1} \|_2^2 + \frac{\lambda}{\alpha} \| s \|_1 \tag{22}$$

Where $s$ and $s_{k+1}$ are the singular values of matrices corresponding to $x$ and $x_{k+1}$ respectively.

It is possible to write (22) in a decoupled fashion,

$$G''_k(s) = (s^{(1)} - s_{k+1}^{(1)})^2 + \frac{\lambda}{\alpha} | s^{(1)} | + \ldots + (s^{(r)} - s_{k+1}^{(r)})^2 + \frac{\lambda}{\alpha} | s^{(r)} | \tag{23}$$

Therefore (23) can be minimized by minimizing each of the terms,

$$f(s^{(i)}) = (s^{(i)} - s_{k+1}^{(i)})^2 + \frac{\lambda}{\alpha} |s^{(i)}| \tag{24}$$

$$\frac{\partial f(s^{(i)})}{\partial s^{(i)}} = 2(s^{(i)} - s_{k+1}^i) + \frac{\lambda}{\alpha} signum(s^{(i)})$$

Setting the derivative to zero and rearranging, we get

$$s_{k+1}^{(i)} = s^{(i)} + \frac{\lambda}{2\alpha} signum(s^{(i)}) \tag{25}$$

The function that minimizes (25) is the following,

$$s^{(i)} = signum(s_{k+1}^i) \max(0, |s_{k+1}^i| - \frac{\lambda}{2\alpha})$$

Written for the entire vector of singular values,

$$s = soft(s_{k+1}, \frac{\lambda}{2\alpha}) = signum(s_{k+1}) \max(0, |s_{k+1}| - \frac{\lambda}{2\alpha}) \tag{26}$$

This is the famous soft Thresholding function used profusely in compressed sensing.

Equations (17) and (26) suggest the Shrinkage Algorithm:

Shrinkage Algorithm

1. $x_k = x_{k-1} + \frac{1}{\alpha} M^T (y - M x_{k-1})$
2. Form the matrix $X_k$ by reshaping $x_k$.
3. SVD: $X_k = U \Sigma V^T$.
4. Soft threshold the singular values: $\hat{\Sigma} = soft(diag(\Sigma), \frac{\lambda}{2\alpha})$
5. $X_{k+1} = U \hat{\Sigma} V^T$. Form $x_{k+1}$ by vectorizing $X_{k+1}$.
6. Update: k=k+1 and return to step 1.

These updates are similar to shrinkage updates proposed in [18, 30], but ours is derived in a more rigorous fashion.

    d.   Rank Minimization via Hard-Thresholding

The rank minimization problem can be expressed as,

$$J(x) = \|y - Mx\|_2^2 + \lambda \text{rank}\|X\| \tag{27}$$

Now using the same MM technique and Landweber iterations as above and following steps till (21), we arrive at the following expression,

$$G_k''(s) = \| s - s_{k+1} \|_2^2 + \frac{\lambda}{\alpha} \| s \|_0 \tag{28}$$

Equation (28) decouples the vectors element-wise, therefore we have

$$G''(s) = (s^{(1)} - s_{k+1}^{(1)})^2 + \frac{\lambda}{\alpha} | s^{(1)} |^0 + ... + (s^{(r)} - s_{k+1}^{(r)})^2 + \frac{\lambda}{\alpha} | s^{(r)} |^0 \tag{29}$$

Since the optimization problem is decoupled, one can proceed to minimize (29) element-wise. We now follow an analysis similar to [30]. To derive the minimum, two cases need to be considered: case 1: $s^{(i)} = 0$ and case 2: $s^{(i)} \neq 0$. The element-wise cost is 0 in the first case. For the second case, the cost is $(s^{(i)})^2 - 2s^{(i)} s_{k+1}^{(i)} + \frac{\lambda}{\alpha}$, the minimum of which is reached when $s^{(i)} = s_{k+1}^{(i)}$.

Comparing the cost on both cases, i.e.

0 if $s^{(i)} = 0$

$-(s_{k+1}^{(i)})^2 + \frac{\lambda}{\alpha}$ if $s^{(i)} = s_{k+1}^{(i)}$

We see that the minimum of each decoupled term in (29) is achieved when,

$$s^{(i)} = \begin{cases} s_{k+1}^{(i)} & \text{when } s_{k+1}^{(i)} > \lambda / 2\alpha \\ 0 & \text{when } s_{k+1}^{(i)} \leq 0 \end{cases} \tag{30}$$

When (30) is applied on the whole vector element-wise, it is popularly called hard-thresholding, and we can write,

$$s = hard(s_{k+1}, \frac{\lambda}{2\alpha}) \tag{31}$$

Based on these equation, we can have a hard-thresholding algorithm similar to the shrinkage algorithm of the previous subsection,

Hard-Thresholding Algorithm

1. $x_k = x_{k-1} + \frac{1}{\alpha} M^T (y - Mx_{k-1})$
2. Form the matrix $X_k$ by reshaping $x_k$.
3. SVD: $X_k = U\Sigma V^T$.
4. Soft threshold the singular values: $\hat{\Sigma} = hard(diag(\Sigma), \frac{\lambda}{2\alpha})$
5. $X_{k+1} = U\hat{\Sigma}V^T$. Form $x_{k+1}$ by vectorizing $X_{k+1}$.
6. Update: k=k+1 and return to step 1.

A similar algorithm was proposed (somewhat heuristically) in [20]. In [20], other intuitive (heuristic) methods have been proposed by a combination of shrinkage and hard-thresholding. However, we cannot arrive at those algorithms following the MM framework.

e. Proposed Non-convex Optimization

The non-convex matrix completion problem needs solving (15). Using the MM procedure and Landweber iterations and following the steps till (21), we arrive at the following,

$$G''(s) = \| s - s_{k+1} \|_2^2 + \frac{\lambda}{\alpha} \| s \|_p, \quad 0 < p < 1 \tag{32}$$

Differentiating (32), we get,

$$\nabla G''(s) = 2s - 2s_{k+1} + \frac{\lambda}{\alpha} Ds \tag{33}$$

where $D = Diag(|s|)^{p-2}$. In practice, the matrix D is evaluated based on the solution of the previous iteration.

Setting the gradient to zero, one gets,

$$s = (I + \frac{\lambda}{2\alpha} D)^{-1} s_{k+1} \tag{34}$$

Both $I$ and $D$ are diagonal, therefore computing the inverse is trivial.

Based on this simple derivation we present the following algorithm:

Non-convex Algorithm

1. $x_k = x_{k-1} + \frac{1}{\alpha} M^T (y - Mx_{k-1})$
2. Form the matrix $X_k$ by reshaping $x_k$.
3. SVD: $X_k = U\Sigma V^T$.
4. Shrink the singular values: $\hat{\Sigma} = (I + \frac{\lambda}{2\alpha} D)^{-1} diag(\Sigma)$
5. $X_{k+1} = U\hat{\Sigma} V^T$. Form $x_{k+1}$ by vectorizing $X_{k+1}$.
6. Update: k=k+1 and return to step 1.

f. Cooling Algorithm

So far, we have been discussing techniques to solve the unconstrained version of the optimization problem (15). However, our actual target is to solve the constrained problem (12). Although the parameters λ and σ are related, the relation is not analytical and is nearly impossible to find for all practical cases. In this paper, we tackle this problem by adopting a cooling technique as employed by [31] in solving the $l_1$-norm minimization problem via iterative soft thresholding.

Cooling Algorithm

Initialize: $x_0 = 0$; $\lambda < \max(M^T x)$

Choose a decrease factor (*DecFac*) for cooling $\lambda$

Outer Loop: While[1] $\|y - Mx\|_2 > \sigma$

Inner Loop: While[2] $\dfrac{J_k - J_{k+1}}{J_k + J_{k+1}} \geq Tol$

Compute objective function for current iterate: $J_k = \|y - Mx_k\|_2^2 + \lambda \|X_k\|_{p*}$

Minimize $J_k$: Depending on the value of p, use either shrinkage (p=1) or hard-thresholding (p=0) or non-convex optimization (0<p<1) to minimize $J_k$.

Compute objective function for next iterate: $J_{k+1} = \|y - Mx_{k+1}\|_2^2 + \lambda \|X_{k+1}\|_{p*}$

End While[2] (inner loop ends)

$\lambda = \lambda \times DecFac$

End While[1] (outer loop ends)

This cooling algorithm is actually based on the smoothness and monotonic decreasing property of the Pareto curve between the objective function $\|X_k\|_{p*}$ and the constraint $\|y - Mx_k\|_2^2$.

## 6. EXPERIMENTAL EVALUATION

In this work we aim at reducing the average energy consumption of the sensor network by sampling and transmitting only a subset of all the sensor nodes at each sampling instant. It is assumed that the sensor nodes lie on a regular rectangular grid. In this work we recast the problem of estimating the values at the unsampled sensor nodes, as a matrix completion problem. Different sampling strategies and reconstruction algorithms has been proposed in this paper.

We do not have access to any real dataset having the required configuration. Therefore we experiment on synthetic data generated by the tool [33]. The tool generates data having characteristics similar to real data as can be seen from the paper associated with the tool [34]. The tool simulates a sensor array of size 64 by 64. The correlation of the sensor network data can be varied. In this work, we generated three types of data having low, medium and high correlation. In this paper, we repeat each experiment 1000 times. The average reconstruction errors of these 1000 iterations are reported.

Some preliminary results using the matrix completion approach to the sensor network problem has been reported in a previous work [35]. A simple random sampling scheme for choosing the sensor samples and a fixed point iteration algorithm proposed in [20] was employed for reconstruction. Moreover, the data acquisition was considered to be noiseless. This work is a non-trivial expansion of [35]. First we consider the realistic scenario where the sampled data is assumed to be corrupted with Gaussian noise. Second, we explore three blue noise sampling schemes apart from the standard random sampling scheme. Third, we derive our own algorithms for convex, non-convex and NP hard matrix completion problem.

First we will experiment on the different blue noise sampling schemes. In tables 2-10, three noise levels (0%, 5% and 10%) are considered, but the reconstruction algorithm is kept the same. We have employed nuclear norm minimization problem via shrinkage to solve the matrix completion problem. Tables 2-4 show results on data having low correlation; tables 5-7 show results on data having medium correlation; tables 8-10 show results on data having high correlation. In all the tables the sampling ratio indicates the ratio of the number of active sensors to the total number of sensors in the array. The tabulated values in all the tables are the normalized mean squared error between the reconstructed data and the actual ground-truth.

Table 2. Low Correlation: Reconstruction Errors for Different Sampling Schemes at 0% Noise

| Sampling Scheme | Sampling Ratio | | | | |
|---|---|---|---|---|---|
| | 10% | 20% | 30% | 40% | 50% |
| Quasi-random | 0.0377 | 0.0167 | 0.0116 | 0.0086 | 0.0071 |
| Quasi-crystal | 0.0415 | 0.0127 | 0.0103 | 0.0085 | 0.0069 |
| Farthest Point | 0.0649 | 0.0221 | 0.0122 | 0.0085 | 0.0070 |
| Random | 0.0742 | 0.0271 | 0.0118 | 0.0105 | 0.0078 |

Table 3. Low Correlation: Reconstruction Errors for Different Sampling Schemes at 5% Noise

| Sampling Scheme | Sampling Ratio | | | | |
|---|---|---|---|---|---|
| | 10% | 20% | 30% | 40% | 50% |
| Quasi-random | 0.1235 | 0.0195 | 0.0128 | 0.0109 | 0.0088 |
| Quasi-crystal | 0.0435 | 0.0191 | 0.0126 | 0.0098 | 0.0076 |
| Farthest Point | 0.4018 | 0.0263 | 0.0186 | 0.0131 | 0.0090 |
| Random | 0.1741 | 0.0297 | 0.0169 | 0.0122 | 0.0097 |

Table 4. Low Correlation: Reconstruction Errors for Different Sampling Schemes at 10% Noise

| Sampling Scheme | Sampling Ratio | | | | |
|---|---|---|---|---|---|
| | 10% | 20% | 30% | 40% | 50% |
| Quasi-random | 0.0422 | 0.0198 | 0.0134 | 0.0117 | 0.0101 |
| Quasi-crystal | 0.0471 | 0.0201 | 0.0131 | 0.0112 | 0.0093 |
| Farthest Point | 0.1695 | 0.0251 | 0.0193 | 0.0130 | 0.0106 |
| Random | 0.0791 | 0.0261 | 0.0162 | 0.0128 | 0.0113 |

Table 5. Medium Correlation: Reconstruction Errors for Different Sampling Schemes at 0% Noise

| Sampling Scheme | Sampling Ratio | | | | |
|---|---|---|---|---|---|
| | 10% | 20% | 30% | 40% | 50% |
| Quasi-random | 0.0350 | 0.0074 | 0.0059 | 0.0046 | 0.0040 |
| Quasi-crystal | 0.0369 | 0.0083 | 0.0058 | 0.0046 | 0.0039 |
| Farthest Point | 0.0510 | 0.0102 | 0.0071 | 0.0048 | 0.0034 |
| Random | 0.0538 | 0.0123 | 0.0068 | 0.0056 | 0.0047 |

Table 6. Medium Correlation: Reconstruction Errors for Different Sampling Schemes at 5% Noise

| Sampling Scheme | Sampling Ratio | | | | |
|---|---|---|---|---|---|
| | 10% | 20% | 30% | 40% | 50% |
| Quasi-random | 0.0376 | 0.0099 | 0.0072 | 0.0061 | 0.0049 |
| Quasi-crystal | 0.0403 | 0.0095 | 0.0062 | 0.0046 | 0.0041 |
| Farthest Point | 0.0519 | 0.0107 | 0.0081 | 0.0059 | 0.0046 |
| Random | 0.0538 | 0.0122 | 0.0075 | 0.0063 | 0.0056 |

Table 7. Medium Correlation: Reconstruction Errors for Different Sampling Schemes at 10% Noise

| Sampling Scheme | Sampling Ratio | | | | |
|---|---|---|---|---|---|
| | 10% | 20% | 30% | 40% | 50% |
| Quasi-random | 0.0350 | 0.0092 | 0.0074 | 0.0063 | 0.0051 |
| Quasi-crystal | 0.0413 | 0.0095 | 0.0065 | 0.0055 | 0.0045 |
| Farthest Point | 0.0507 | 0.0111 | 0.0086 | 0.0062 | 0.0046 |
| Random | 0.0511 | 0.0145 | 0.0076 | 0.0065 | 0.0060 |

Table 8. High Correlation: Reconstruction Errors for Different Sampling Schemes at 0% Noise

| Sampling Scheme | Sampling Ratio | | | | |
|---|---|---|---|---|---|
| | 10% | 20% | 30% | 40% | 50% |
| Quasi-random | 0.1312 | 0.0059 | 0.0045 | 0.0038 | 0.0029 |
| Quasi-crystal | 0.0287 | 0.0060 | 0.0046 | 0.0038 | 0.0029 |
| Farthest Point | 0.0451 | 0.0075 | 0.0057 | 0.0042 | 0.0032 |
| Random | 0.0536 | 0.0082 | 0.0053 | 0.0045 | 0.0035 |

Table 9. High Correlation: Reconstruction Errors for Different Sampling Schemes at 5% Noise

| Sampling Scheme | Sampling Ratio | | | | |
|---|---|---|---|---|---|
| | 10% | 20% | 30% | 40% | 50% |
| Quasi-random | 0.0320 | 0.0079 | 0.0052 | 0.0040 | 0.0034 |
| Quasi-crystal | 0.0135 | 0.0064 | 0.0052 | 0.0038 | 0.0031 |
| Farthest Point | 0.0509 | 0.0071 | 0.0062 | 0.0040 | 0.0033 |
| Random | 0.0524 | 0.0140 | 0.0060 | 0.0046 | 0.0038 |

Table 10. High Correlation: Reconstruction Errors for Different Sampling Schemes at 10% Noise

| Sampling Scheme | Sampling Ratio | | | | |
|---|---|---|---|---|---|
| | 10% | 20% | 30% | 40% | 50% |
| Quasi-random | 0. 1346 | 0.0082 | 0.0057 | 0.0045 | 0.0034 |
| Quasi-crystal | 0.0172 | 0.0073 | 0.0051 | 0.0040 | 0.0031 |
| Farthest Point | 0.0479 | 0.0084 | 0.0070 | 0.0047 | 0.0033 |
| Random | 0.0559 | 0.0102 | 0.0063 | 0.0044 | 0.0039 |

From these tables, we find that our blue noise sampling schemes – Quasi-random, Quasi-crystal and Farthest Point sampling give far superior results compared to random sampling in terms of reconstruction accuracy. Of the three Quasi-crystal sampling is the most stable. Therefore for the second part of our experiments we fix the sampling scheme to Quasi-crystal sampling and see how the reconstruction error varies when different reconstruction algorithms – Shrinkage (nuclear norm minimization), Hard-Thresholding (rank minimization) and Non-convex optimization are employed. The value of p for the non-convex algorithm is fixed at 0.8. Tables 11-19 show the reconstruction results for different reconstruction algorithms at three different noise levels (0%, 5% and 10%).

Table 11. Low Correlation: Reconstruction Errors for Different Algorithms at 0% Noise

| Algorithm | Sampling Ratio | | | | |
|---|---|---|---|---|---|
| | 10% | 20% | 30% | 40% | 50% |
| Shrinkage | 0.0415 | 0.0127 | 0.0103 | 0.0085 | 0.0069 |
| Hard-Thresholding | 0.1327 | 0.0509 | 0.0127 | 0.0080 | 0.0067 |
| Non-Convex Optimization | 0.0298 | 0.0113 | 0.0098 | 0.0080 | 0.0067 |

Table 12. Low Correlation: Reconstruction Errors for Different Algorithms at 5% Noise

| Algorithm | Sampling Ratio | | | | |
|---|---|---|---|---|---|
| | 10% | 20% | 30% | 40% | 50% |
| Shrinkage | 0.0435 | 0.0191 | 0.0126 | 0.0098 | 0.0076 |
| Hard-Thresholding | 0.2076 | 0.0678 | 0.0135 | 0.0096 | 0.0082 |
| Non-Convex Optimization | 0.0364 | 0.0188 | 0.0106 | 0.0093 | 0.0076 |

Table 13. Low Correlation: Reconstruction Errors for Different Algorithms at 10% Noise

| Algorithm | Sampling Ratio | | | | |
|---|---|---|---|---|---|
| | 10% | 20% | 30% | 40% | 50% |
| Shrinkage | 0.0471 | 0.0201 | 0.0131 | 0.0112 | 0.0093 |
| Hard-Thresholding | 0.2174 | 0.0235 | 0.0157 | 0.0100 | 0.0081 |
| Non-Convex Optimization | 0.0385 | 0.0197 | 0.0119 | 0.0097 | 0.0080 |

Table 14. Medium Correlation: Reconstruction Errors for Different Algorithms at 0% Noise

| Algorithm | Sampling Ratio | | | | |
|---|---|---|---|---|---|
| | 10% | 20% | 30% | 40% | 50% |
| Shrinkage | 0.0369 | 0.0083 | 0.0058 | 0.0046 | 0.0039 |
| Hard-Thresholding | 0.0786 | 0.0095 | 0.0060 | 0.0042 | 0.0036 |
| Non-Convex Optimization | 0.0284 | 0.0076 | 0.0056 | 0.0042 | 0.0035 |

Table 15. Medium Correlation: Reconstruction Errors for Different Algorithms at 5% Noise

| Algorithm | Sampling Ratio | | | | |
|---|---|---|---|---|---|
| | 10% | 20% | 30% | 40% | 50% |
| Shrinkage | 0.0403 | 0.0095 | 0.0062 | 0.0046 | 0.0041 |
| Hard-Thresholding | 0.0892 | 0.0108 | 0.0067 | 0.0046 | 0.0041 |
| Non-Convex Optimization | 0.0309 | 0.0089 | 0.0060 | 0.0045 | 0.0041 |

Table 16. Medium Correlation: Reconstruction Errors for Different Algorithms at 5% Noise

| Algorithm | Sampling Ratio | | | | |
|---|---|---|---|---|---|
| | 10% | 20% | 30% | 40% | 50% |
| Shrinkage | 0.0413 | 0.0095 | 0.0065 | 0.0055 | 0.0045 |
| Hard-Thresholding | 0.1058 | 0.0112 | 0.0071 | 0.0052 | 0.0043 |
| Non-Convex Optimization | 0.0317 | 0.0101 | 0.0065 | 0.0052 | 0.0044 |

Table 17. High Correlation: Reconstruction Errors for Different Algorithms at 0% Noise

| Algorithm | Sampling Ratio | | | | |
|---|---|---|---|---|---|
| | 10% | 20% | 30% | 40% | 50% |
| Shrinkage | 0.0287 | 0.0060 | 0.0046 | 0.0038 | 0.0029 |
| Hard-Thresholding | 0.0598 | 0.0078 | 0.0050 | 0.0039 | 0.0029 |
| Non-Convex Optimization | 0.0245 | 0.0057 | 0.0044 | 0.0036 | 0.0028 |

Table 18. High Correlation: Reconstruction Errors for Different Algorithms at 5% Noise

| Algorithm | Sampling Ratio | | | | |
|---|---|---|---|---|---|
| | 10% | 20% | 30% | 40% | 50% |
| Shrinkage | 0.0135 | 0.0064 | 0.0052 | 0.0038 | 0.0031 |
| Hard-Thresholding | 0.0609 | 0.0102 | 0.0059 | 0.0038 | 0.0030 |
| Non-Convex | 0.0130 | 0.0060 | 0.0050 | 0.0038 | 0.0031 |
| Optimization | | | | | |

Table 19. High Correlation: Reconstruction Errors for Different Algorithms at 10% Noise

| Algorithm | Sampling Ratio | | | | |
|---|---|---|---|---|---|
| | 10% | 20% | 30% | 40% | 50% |
| Shrinkage | 0.0172 | 0.0073 | 0.0051 | 0.0040 | 0.0031 |
| Hard-Thresholding | 0.0647 | 0.0110 | 0.0060 | 0.0041 | 0.0031 |
| Non-Convex Optimization | 0.0155 | 0.0064 | 0.0051 | 0.0040 | 0.0032 |

From tables 11-19 we can infer that the non-convex optimization almost always give the best reconstruction results. The results from the shrinkage algorithm close follow the ones from non-convex optimization. Hard-thresholding gives good results when the sampling ratio is higher; at low sampling ratios the reconstruction is very unstable.

The experimental results follow our intuition. As the correlation in the data increases, the reconstruction error improves when the sampling ratio is kept constant. Matrix completion is basically an interpolation method assuming that the rank of the matrix is low. When the correlation in the data is high, any interpolation method will yield better results; the results from matrix completion are no different. Also we observe that when the noise in the data increases, the reconstruction error degrades slightly. The same phenomenon is observed in any other denoising/reconstruction problem. Higher the noise, the more difficult it is to extract the original signal.

7. CONCLUSION

In this work, we propose a fundamentally new approach to reduce the energy consumption in wireless sensor networks. At each sampling instant only a subset of all the sensors are active. The active sensors only acquire and transmit the data to the sink. The sink employs a smart reconstruction algorithm to estimate the sample values at all (sampled as well as unsampled) the sensor nodes. At the next sampling instant a different set of sensor nodes will be active. Thus it is guaranteed that the life of the sensor is increased.

The proposed method has several advantages. First, the scheme is data-independent, therefore complex model driven sampling strategy is required to chose the positions of the sensors to be sampled. Second, it reduces the number of sampling and transmissions to be made, thereby reducing the operations of the two most power hungry modules of the sensor node. The third advantage of the proposed approach is that it can work with most of previous power reduction strategies of duty cycling [36] and data driven prediction [37].

It is assumed that the sensors are located on a regular rectangular grid. Therefore the problem of estimating the values at different sensor nodes could be cast as a matrix completion problem. In this work we empirically extend the findings of recent theoretical works in matrix completion. First, we show that more sophisticated blue noise sampling schemes like Quasi-crystal sampling, Farthest Point sampling and

Quasi-random sampling give superior results compared to simple random sampling. Till date, all the matrix completion theory has only been proved for random sampling strategies. Our experimental results will motivate theoretical studies into such blue noise sampling schemes as part of matrix completion theory. The second contribution of this paper is in deriving the matrix completion algorithms in a rigorous optimization framework. We derived a shrinkage algorithm for solving matrix completion via nuclear norm minimization, a hard-thresholding algorithm for solving matrix completion via rank minimization and a non-convex optimization algorithm for matrix completion based on minimizing the $l_p$-norm of the singular values. The non-convex algorithm gives the best reconstruction results.